\documentclass[showpacs,preprintnumbers,amsmath,amssymb,twocolumn]{revtex4}
\usepackage{amsmath,amsfonts,latexsym,amssymb,graphicx,graphics,epsfig,subfigure,color,makeidx}
\usepackage{xcolor,diagbox}
\usepackage{multirow}
\usepackage[colorlinks,linkcolor=blue,anchorcolor=blue,citecolor=green,urlcolor=blue]{hyperref}
\usepackage{lipsum}
\usepackage{appendix}

\usepackage[latin1]{inputenc}
\newcommand {\nn}    {\nonumber}
\newcommand {\fc}    {\frac}
\newcommand {\be}    {\begin{equation}}
\newcommand {\ee}    {\end{equation}}
\newcommand {\beq}   {\begin{eqnarray}}
\newcommand {\eeq}   {\end{eqnarray}}
\newcommand {\pd}    {\partial}
\newcommand {\lt}    {\left}
\newcommand {\rt}    {\right}

\begin{document}

\title{Gravito-Electromagnetic coupled perturbations and quasinormal modes of a charged black hole with scalar hair}

\author{Wen-Di Guo$^{ab}$\footnote{Wen-Di Guo and Qin Tan are co-first authors of this paper.}}
\author{Qin Tan$^{ab}$}
\author{Yu-Xiao Liu$^{ab}$\footnote{liuyx@lzu.edu.cn, corresponding author}}

\affiliation{$^{a}$Lanzhou Center for Theoretical Physics, Key Laboratory of Theoretical Physics of Gansu Province,
School of Physical Science and Technology, Lanzhou University, Lanzhou 730000, People's Republic of China\\
             $^{b}$Institute of Theoretical Physics $\&$ Research Center of Gravitation, Lanzhou University, Lanzhou 730000, People's Republic of China}

\begin{abstract}
From the quantum point of view, singularity should not exist. Recently, Bah and Heidmann constructed a five-dimensional singularity free topology star/black hole [Phys. Rev. Lett. 126, 151101 (2021)]. By integrating the extra dimension, a four-dimensional static spherically symmetric black hole with a magnetic charge and scalar hair can be obtained. In this paper, we study the quasinormal modes (QNMs) of the magnetic field and gravitational field on the background of this four-dimensional charged black hole with scalar hair. The odd parity of the gravitational perturbations couples with the even parity of the magnetic field perturbations. Two coupled second-order derivative equations are obtained. Using the matrix-valued direct integration method and the matrix-valued continued fraction method, we obtain the fundamental QNM frequencies numerically. The effect of the magnetic charge on the QNMs is studied. The differences of the frequencies of the fundamental QNMs between the charged black hole with scalar hair and the Reissner-Norstr\"{o}m black hole are very small for the angular number $l=2$.
%However, some new interesting results are found for higher angular number.
\end{abstract}

%\begin{keyword}
%%% keywords here, in the form: keyword \sep keyword
%Perturbation, $f(T)$ brane world, stability
%%% MSC codes here, in the form: \MSC code \sep code
%%% or \MSC[2008] code \sep code (2000 is the default)
%Mimetic theory, $f(T)$ theory, Extra dimensions, Brane-word, Gravity localization
%\end{keyword}

%\pacs{04.50.-h, 11.27.+d }

\maketitle

\section{Introduction}

Black hole physics has entered a new era since the detection of the gravitational waves from a binary black hole merger by Laser Interferometer Gravitational-Wave Observatory (LIGO) and Virgo~\cite{Collaboration2016} and the first picture of a supermassive black hole at the center of galaxy M87 photographed by the Event Horizon Telescope (EHT)~\cite{Collaboration2019,Collaboration2019a,Collaboration2019b,Collaboration2019c,Collaboration2019d,Collaboration2019e}. Recently, the picture of the black hole in our Milky way was also taken by EHT~\cite{Collaboration2022,Collaboration2022a,Collaboration2022b,Collaboration2022c,Collaboration2022d,Collaboration2022e}. These breakthroughs provide us with more possibilities to test some fundamental physical problems, for example, the singularity problem in mathematical aspect~\cite{Berti2015,Cardoso2016}. Usually, a spacetime singularity is located at the center of a black hole. However, from the quantum aspect, spacetime should not be singular. In order to mimic black holes classically, some ultra-compact objects have been constructed, such as, gravastars~\cite{Mazur2001}, boson stars~\cite{Schunck2008}, and wormholes~\cite{Solodukhin2005,Dai:2019mse,Simonetti:2020ivl,Bambi:2021qfo}. For more details, see the review ~\cite{Cardoso2019} and references therein. But usually they need some exotic matters and the UV origin is unclear.  From the top-down point view, string theory is regarded as the candidate that can unify quantum theory and gravity. Some horizonless models constructed from string theory, such as fuzz balls~\cite{Gibbons2013}, are similar to black holes up to the Planck scale, and they have smooth microstate geometries. However, a lot of degrees of freedom in supergravity are needed, and the astrophysical observations of these horizonless models are difficult~\cite{Bena2020,Bena2020a,Bena2020b}. Recently, a five-dimensional nonsingular topological star/black hole model was proposed based on a five-dimensional Einstein-Maxwell theory~\cite{Bah2020,Bah2020a}. The spacetime in this model has advantages both in microstate (smooth geometry) and macrostate geometries (similar to classical black holes). So it is interesting to study their astrophysical observations. Last year, Lim studied the motion of a charged particle in this nonsingular topological star/black hole model~\cite{Lim2021}. The thermodynamic stability of the solutions has been carefully analysed in Ref.~\cite{Bah:2021irr}. Integrating the extra dimension, a four-dimensional Einstein-Maxwell-Dilaton theory can be obtained, and a static spherically symmetric solution was solved in this background~\cite{Bena2020a,Bena2020b}. Shadows of this black hole were studied in Ref.~\cite{Guo2022}. In this paper, we will study the quasinormal modes (QNMs) of this model.

As the characteristic modes of a dissipative system, QNMs play important roles in a lot of aspects of our world. Due to the presence of the event horizon, black holes are natural dissipative systems. For a binary black hole merger system, there are three stages: inspiral, merger, and ringdown. In the ringdown stage, the gravitational waves are regarded as a superposition of QNMs~\cite{Berti2007}. Compared with the normal modes, the eigenfunctions of QNMs generally do not form a complete set, and they are not normalizable~\cite{Nollert1998}. The frequencies of QNMs are complex, and the imaginary parts are related to the decay timescale of the perturbation. One can use the QNMs to infer the mass and angular momentum of a black hole~\cite{Echeverria:1989hg} and to test the validity of the no-hair theorem~\cite{Berti:2005ys,Berti:2007zu,Isi:2019aib}. The echoes in the ringdown signals can be used to distinguish the black hole from the ultra compact objects~\cite{Cardoso2016,Cardoso2019,Cardoso:2017cqb}. Recently, the pseudospectrum of gravitational physics showed that the QNM spectrum is unstable for the fundamental mode and the overtone modes~\cite{Jaramillo:2020tuu,Cheung:2021bol}. Besides, the properties of QNMs can also be used to constrain modified gravity theories~\cite{Wang:2004bv,Blazquez-Salcedo:2016enn,Franciolini:2018uyq,Aragon:2020xtm,Liu:2020qia,Karakasis:2021tqx,Cano:2021myl,Gonzalez:2022upu,Zhao:2022gxl}.
The stability under perturbations of the background spacetime can also be partly revealed from the QNM frequencies~\cite{Ishibashi:2003ap,Chowdhury:2022zqg}. Except for black hole physics, QNMs are also very useful in other dissipative systems, such as leaky resonant cavities~\cite{Kristensen2015}, and brane world theories~\cite{Seahra2005,Seahra2005a,Tan2022}. So, QNMs have been studied widely~\cite{Cai:2015fia,Cardoso:2019mqo,McManus:2019ulj,Cardoso:2020nst,Guo:2021enm}.

In this paper, we are interested in the QNMs of the four-dimensional spherically symmetric Bah-Heidmann black hole with a magnetic charge. The organization of this paper is as follows. In Sec.~\ref{themodel}, we briefly review the Bah-Heidmann black hole and the Kaluza-Klein (KK) reduction. In Sec.~\ref{pereqs}, we study the linear perturbation of the electromagnetic field and gravitational field. Separating radial part of the perturbed fields from the angular part, we derive the perturbation equations. In Sec.~\ref{QNM}, we compute the quasinormal frequencies (QNFs) using the matrix-valued direct integration method. Finally, we give our conclusions in Sec.~\ref{conclusion}.

\section{The charged black hole with scalar hair}\label{themodel}

In this section we briefly review the black hole/topological star model proposed by Bah and Heidmann~\cite{Bah2020,Bah2020a}. We start from a five-dimensional Einstein-Maxwell theory. The action is
\begin{equation}
S=\int d^5x\sqrt{-\hat{g}}\left(\fc{1}{16\pi G_5}\hat{R}-\fc{1}{16\pi}\hat{F}^{MN}\hat{F}_{MN}\right),\label{action5}
\end{equation}
where $\hat{F}_{MN}$ is the electromagnetic field tensor and $G_{5}$ is the five-dimensional gravitational constant. The quantities with hat denote that they are constructed in the five-dimensional spacetime. The capital Latin letters $M, N...$ denote the five-dimensional coordinates. The metric can be assumed as~\cite{Stotyn2011}
\begin{eqnarray}
ds^2&=&-f_S(r)dt^2+f_B(r)dy^2+\frac{1}{f_S(r)f_B(r)}dr^2\nn\\
&+&r^2d\theta^2+r^2\sin^2\theta d\phi^2. \label{metric_five}
\end{eqnarray}
The extra dimension, denoted by the coordinate $y$, is a warped circle with radius $R_y$.
The field strength with a magnetic flux is
\be
\hat{F}=P\sin\theta d\theta\wedge d\phi.\label{field_strength}
\ee
The solution with double Wick rotation symmetry is~\cite{Stotyn2011}
\beq
 f_{B}(r)&=&1-\frac{r_B}{r},  \nonumber \\
 f_{S}(r)&=&1-\frac{r_S}{r},  \\
 P&=&\pm\frac{1}{G_{5}}\sqrt{3r_S r_B}.\nonumber
\eeq
That is to say, the metric~\eqref{metric_five} is invariant under rotation ($t$, $y$, $r_S$, $r_B$) $\rightarrow$ ($iy$, $it$, $r_B$, $r_S$). There are two coordinate singularities located at $r=r_S$ (corresponding to a horizon) and $r=r_B$ (corresponding to a degeneracy of the $y$-circle). Bah and Heidmann found that, after some coordinate transformations, a smooth bubble locates at $r=r_B$~\cite{Bah2020,Bah2020a}. This provides an end of the spacetime. For $r_S\geq r_{B}$, the bubble is hidden behind the horizon and the metric~\eqref{metric_five} describes a black string. For $r_S<r_{B}$, the spacetime ends at the bubble before reaching the horizon and the metric~\eqref{metric_five} describes a topological star~\cite{Bah2020,Bah2020a}.

We can integrate the extra dimension $y$ (this process is called  Kaluza-Klein reduction). Then, a four-dimensional  Einstein-Maxwell-dilaton theory is obtained from the five-dimensional Einstein-Maxwell theory
\beq
S_4&=&\int d^4x\sqrt{-g}\Big(\fc{1}{16\pi G_4}R_4-\fc{3}{8\pi G_4}g^{\mu\nu}\pd_{\mu}\Phi\pd_{\nu}\Phi\nn\\
&-& \fc{e^{-2\Phi}}{16\pi e^2}F_{\mu\nu}F^{\mu\nu}\Big)\label{action4},
\eeq
where $e^2\equiv\fc{1}{2\pi R_y}$ and $\Phi$ is a dilaton field. The Greek letters $\mu, \nu,...$ denote the four-dimensional coordinates. Here, $g_{\mu\nu}$ and $F_{\mu\nu}$ are the four-dimensional metric~\eqref{metric_four} and the electromagnetic field strength, respectively. The four-dimensional Ricci scalar $R_4$ is determined by the metric $g_{\mu\nu}$, and the four-dimensional gravitational constant is defined as
\be
 G_4=e^2 G_5.
\ee
Varying the action~\eqref{action4} with respect to the scalar field $\Phi$, the vector potential $A_{\mu}$, and the metric $g_{\mu \nu}$, we obtain the field equations
\beq
\fc{6}{G_4}\Box\Phi+\fc{e^{-2\Phi}}{e^2}F_{\mu\nu}F^{\mu\nu}&=&0,\label{EFphi}\\
\nabla^{\mu}F_{\mu\nu}&=&0,\label{EFA}\\
R_{\mu\nu}-\fc{1}{2}g_{\mu\nu}&=&8\pi G_4 T_{\mu\nu},\label{EFg}
\eeq
where $\Box$ is the four-dimensional D'Alembert operator, $T_{\mu\nu}=T^\text{s}_{\mu\nu}+T^\text{m}_{\mu\nu}$ is the energy momentum tensor containing the contributions of the scalar field and the magnetic field:
\beq
T^\text{s}_{\mu\nu}&=&\fc{3}{4\pi G_4}\nabla_{\mu}\Phi\nabla_{\nu}\Phi-\fc{3}{8\pi G_4}g_{\mu\nu}\Box\Phi,\\
T^\text{m}_{\mu\nu}&=&\fc{e^{-2\Phi}}{4\pi e^2}F_{\mu\alpha}F_{\nu}^{\alpha}-\fc{e^{-2\Phi}}{16\pi e^2}g_{\mu\nu}F_{\alpha\beta}F^{\alpha\beta}.
\eeq
The dilaton field $\Phi$ can be obtained as
\be
e^{2\Phi}=f_B^{-1/2}\label{solutionphi}.
\ee
We can solve the vector potential corresponding to the magnetic field as
\be
A_{\mu}=(0,0,0,-\fc{e}{2}\sqrt{\fc{3r_B r_S}{G_4}}\cos\theta)\label{solutionnF}.
\ee
Thus, the field strength reads as
\be
F_{\mu\nu}=\lt[\begin{array}{cccc}
0&0&0&0\\
0&0&0&0\\
0&0&0&\fc{e}{2}\sqrt{\fc{3r_B r_S}{G_4}}\sin\theta\\
0&0&-\fc{e}{2}\sqrt{\fc{3r_B r_S}{G_4}}\sin\theta&0
\end{array}\rt].
\ee
The four-dimensional metric is
\beq
ds_4^2=f_B^{\frac{1}{2}}\lt(-f_Sdt^2+\frac{dr^2}{f_Bf_S}+r^2d\theta^2+r^2\sin^2\theta d\phi^2\rt).\nn \\ \label{metric_four}
\eeq
Note that, when $r_B=0$, this metric recovers to the Schwarzschild one.

The parameters $r_S$ and $r_B$ are related to the four-dimensional Arnowitt-Deser-Misner mass $M$ and the magnetic charge $Q_\text{m}$ as
\beq
M&=&\left(\fc{2r_S+r_B}{4 G_4}\right),\\
Q_\text{m}&=&\fc{1}{2}\sqrt{\fc{3r_B r_S}{G_4}}.
\eeq
On the other hand, for each $M$ and $Q_\text{m}$, which are physical parameters, there are two solutions of ($r_S$,$r_B$)
\beq
r_S^{(1)}&=&2G_4(M-M_{\triangle}),~~~r_B^{(1)}=G_4(M+M_{\triangle});\label{MQ1}\\
r_S^{(2)}&=&G_4(M+M_{\triangle}),~~~~r_B^{(2)}=2G_4(M-M_{\triangle}),\label{MQ2}
\eeq
where
\be
M_{\triangle}^2=M^2-\left(\fc{\sqrt{2} Q_\text{m}}{\sqrt{3 G_4 }}\right)^2.\label{Mdelta}
\ee
Note that, in four-dimensional spacetime, when $r<r_B$, $f_B^{1/2}$ becomes imaginary. So, $r=r_B$ is the end of the spacetime. This is consistent with the result in five-dimensional spacetime~\cite{Bah2020,Bah2020a}.  Usually, a black string scenario has the Gregory-Laflamme instability~\cite{Gregory:1993vy}. However, compact extra dimensions leading to a discrete KK mass spectrum makes it possible to avoid the Gregory-Laflamme instability. Stotyn and Mann demonstrated that, the solution~\eqref{MQ1} is unstable under perturbation, while, when $R_y>\fc{4\sqrt{3}}{3}Q_{\text{m}}$, the solution~\eqref{MQ2} is stable. That is to say, the solution~\eqref{MQ2} does not have the Gregory-Laflamme instability. Actually, the spacetime at $r=r_B$ is singular in four-dimensional spacetime. When $r_B\geq r_S$ the metric~\eqref{metric_four} corresponds to a naked singularity, and when $r_B<r_S$ the metric~\eqref{metric_four} corresponds to a black hole, which is named as a charged black hole with scalar hair. In this paper, we will only focus on the case $r_B<r_S$, i.e., the charged black hole with scalar hair.

\section{Perturbation equations}\label{pereqs}

With the background solution \eqref{solutionphi},~\eqref{metric_four}, and \eqref{solutionnF}, we can derive the equations of motion for the perturbations. The perturbed scalar field, vector potential, and metric field can be written as
\beq
\Phi&=&\bar{\Phi}+\varphi,\\
A_{\mu}&=&\bar{A}_{\mu}+a_{\mu},\\
g_{\mu\nu}&=&\bar{g}_{\mu\nu}+h_{\mu\nu},
\eeq
where the quantities with a bar represent the background fields, $\varphi$, $a_{\mu}$, and $h_{\mu\nu}$ denote the corresponding perturbations. Because the background spacetime is spherically symmetric, the perturbations can be divided into three parts based on their transformations under rotations on the 2-sphere: scalars, two-dimensional vectors, and two-dimensional tensors. The spherical harmonic function $Y_{l,m}(\theta,\phi)$ behaves as a scalar under rotations, so it is the scalar base. The two-dimensional vector and tensor bases are introduced as follows~\cite{Wheeler,Ruffini,Edmonds,Regge,Chandrasekhar}
\beq
(V^1_{l,m})_a&=&\pd_a Y_{l,m}(\theta,\phi), \\
(V^2_{l,m})_a&=&\gamma^{bc}\epsilon_{ac}\pd_b Y_{l,m}(\theta,\phi),
\eeq
for the vector part, and
\beq
(T^1_{l,m})_{ab}&=&(Y_{l,m})_{;ab}, \\
(T^2_{l,m})_{ab}&=&Y_{l,m}\gamma_{ab},\\
(T^3_{l,m})_{ab}&=&\fc{1}{2}\lt[\epsilon_a^c(Y_{l,m})_{;cb}+\epsilon_b^c(Y_{l,m})_{;ca}\rt],
\eeq
for the tensor part. Here, the Latin letters $a, b, c$ denote the angular coordinates $\theta$ and $\phi$, $\gamma$ is the induced metric on the 2-sphere with radius $1$, and $\epsilon$ is the totally antisymmetric tensor in two dimensions. The semicolon denotes the covariant derivative on the 2-sphere.

The above quantities behave differently under the space inversion, i.e., $(\theta,\phi)\rightarrow(\pi-\theta,\pi+\phi)$. A quantity is called even or polar, if it acquires a factor $(-1)^l$ under space inversion. A quantity is called odd or axial, if it acquires a factor $(-1)^{l+1}$ under space inversion. So the above quantities can be divided into two classes, the even parts $V^1_{l,m}, T^1_{l,m}, T^2_{l,m}$, and the odd parts $V^2_{l,m}, T^3_{l,m}$. Note that, the spherical harmonic function $Y_{l,m}(\theta,\phi)$ is even-parity. Usually the gravitational and electromagnetic perturbations will mix, for example RN black hole. But the even-parity and odd-parity perturbations usually do not mix, the RN black hole with electric charge do not mix the polar and axial contributions. Only the even-parity (or odd-parity) perturbations of the gravitational and electromagnetic parts mix. However, we can see from Eqs.~\eqref{solutionphi},~\eqref{metric_four}, and~\eqref{solutionnF} that the background scalar field and metric field are even-parity and the background vector potential is odd-parity. So we expect that the scalar perturbation and  even-parity parts of the metric perturbations couple to the odd-parity parts of the electromagnetic perturbations to the linear order (type-I coupling). And the odd-parity parts of the metric perturbations couple to the even-parity parts of the electromagnetic perturbations to the linear order (type-II coupling). Note that, the scalar perturbation only contains the even part. Actually, these coupled perturbation equations have been studied in Refs.~\cite{Nomura:2020tpc,Meng:2022oxg}. In this paper, we study the type-II coupling perturbations.

Based on the principle of general covariance, the theory should keep covariant under an infinitesimal coordinate transformation. Thus, we can choose a specific  gauge to simplify the problem. In the Regge-Wheeler gauge~\cite{Regge}, the odd parts of the perturbation $h_{\mu\nu}$ can be written as
\be
h_{\mu\nu}=\sum_{l}e^{-i\omega t}\lt[\begin{array}{cccc}
0&0&0&h_0\\
0&0&0&h_1\\
0&0&0&0\\
*&*&0&0
\end{array}\rt]\sin\theta\pd_{\theta}Y_{l,0}(\theta).\label{hmunu}
\ee
The magnetic field also has a gauge freedom. Following Ref.~\cite{Zerilli:1973}, we denote
\be
\tilde{f}_{\mu\nu}=\partial_{\mu}a_{\nu}-\partial_{\nu}a_{\mu}\label{fmunu},
\ee
the even parts of the perturbation $\tilde{f}_{\mu\nu}$ can be written as
\be
\tilde{f}_{\mu\nu}=\sum_{l}e^{-i\omega t}\lt[\begin{array}{cccc}
0&f_{01}&f_{02}&0\\
*&0&f_{12}&0\\
0&*&0&0\\
0&*&0&0
\end{array}\rt]\sin\theta\pd_{\theta}Y_{l,0}(\theta).\label{ffmunu}
\ee
Note that, we have chosen $m=0$ for simplicity, because the perturbation equations do not depend on the value of $m$~\cite{Regge}. The asterisks denote elements obtained by symmetry. The functions $h_0, h_1, f_{01}, f_{02},$ and $f_{12}$ only depend on the coordinate $r$. The perturbation of the vector potential can be expanded as
\beq
a_t&=&-\sum_{l}e^{-i\omega t}f_{02}Y_{l,0},\\
a_r&=&-\sum_{l}e^{-i\omega t}f_{12}Y_{l,0},\\
a_{\theta}&=&0,\\
a_{\phi}&=&0.
\eeq
The field strength $f_{01}$ can be derived from Eq.~\eqref{fmunu} as
\be
f_{01}=\pd_r f_{02}+i\omega f_{12}.\label{eqf01}
\ee
Substituting Eqs.~\eqref{hmunu} and \eqref{ffmunu} into the equations of motion \eqref{EFA} and \eqref{EFg}, after some algebra calculations we can obtain the following master perturbation equations
\beq
\fc{d^2\psi_\text{g}}{dr_*^2}+(\omega^2-V_{11})\psi_\text{g}-V_{12}\psi_\text{m}&=&0,\label{mastereq1}\\
\fc{d^2\psi_\text{m}}{dr_*^2}+(\omega^2-V_{22})\psi_\text{m}-V_{21}\psi_\text{g}&=&0,\label{mastereq2}
\eeq
where
\beq
\psi_\text{g}&\equiv& f_B^{1/4}f_S\fc{1}{r}h_1,\\
\psi_\text{m}&\equiv& \sqrt{f_B}r^2f_{01},
\eeq
$r_*$ is the tortoise coordinate defined as
\be
dr_*=\fc{1}{\sqrt{f_B}f_S}dr\label{tort},
\ee
and
\beq
V_{11}&=&f_S\lt[\fc{l(l+1)}{r^2}-\fc{3(r_B^2(13r_S-9r)+16r_S r^2)}{16f_B r^5}\rt]\nn\\
&+&f_S\fc{3r_B (2r-7r_S)}{4f_B r^4}, \label{V11}\\
V_{12}&=&-\fc{2i f_S f_B^{1/4}}{el(l+1)r^3}\sqrt{3r_B r_S G_4}\omega , \label{V12} \\
V_{21}&=&\fc{i\sqrt{3 r_B r_S}e f_S}{2\sqrt{G_4}\omega f_B^{1/4}r^3}(l-1)l(l+1)(l+2), \label{V21}\\
V_{22}&=&f_S\lt[\fc{3r_B r_S}{r^4}+\fc{l(l+1)}{r^2}\rt]\label{V22}.
\eeq
The details of deriving the master equation ~\eqref{mastereq1} and \eqref{mastereq2} are shown in Appendix~\ref{appendix}.

Note that, when the magnetic charge $Q_\text{m}$ vanishes, or $r_B$ approaches to zero, the gravitational perturbation $\psi_\text{g}$ and the magnetic field perturbation $\psi_\text{m}$ will decouple. Furthermore, the potential $V_{11}$ will reduce to the potential for the gravitational perturbation of the Schwarzschild black hole. Besides, the parameters $e$ and $G_4$ do not affect the quasinormal modes. To see this, we can redefine
\beq
 \tilde\psi_\text{m}\equiv \fc{\sqrt{G_4}}{e}\psi_\text{m}
\eeq
to eliminate the parameters $e$ and $G_4$ in Eqs.~\eqref{mastereq1} and \eqref{mastereq2}. The corresponding potentials are
\beq
\tilde{V}_{12}&=&-\fc{2i f_S f_B^{1/4}}{l(l+1)r^3}\sqrt{3r_B r_S}\omega , \label{V12ok}\\
\tilde{V}_{21}&=&\fc{i\sqrt{3 r_B r_S}f_S}{2\omega f_B^{1/4}r^3}(l-1)l(l+1)(l+2).\label{V21ok}
\eeq
In the following, we use the redefined quantities but omit the tilde above them.

\section{Quasinormal modes}\label{QNM}

In this section we will solve the master perturbation equations \eqref{mastereq1} and \eqref{mastereq2} to obtain the frequencies of the QNMs. We focus on the QNMs of the solution~\eqref{MQ2}, because it is free of the Gregory-Laflamme instability. We know from Eq.~\eqref{Mdelta} that the range of the magnetic charge $Q_\text{m}$ is $[0,\sqrt{\fc{3}{2}G_4}M]$. Compared with the range of the electric charge of the Reissner-Norstr\"{o}m (RN) black hole $[0,\sqrt{G_4} M]$, the range of the magnetic charge is larger than that of the RN black hole electric charge. Note that, we only study the charged black hole with scalar hair, that is, $r_B<r_S$. In this situation, the range of the magnetic charge $Q_\text{m}$ is $[0,2\sqrt{\fc{G_4}{3}}M]$. This range is still larger than that of the RN black hole electric charge.

The perturbation equations \eqref{mastereq1} and \eqref{mastereq2} are coupled and can be rewritten into a compact form
\be
\fc{d^2\mathbf{Y}}{dr_*^2}+(\omega^2-\mathbf{V})\mathbf{Y}=0,
\ee
where
\begin{displaymath}
\mathbf{Y}=
\left( \begin{array}{c}
\psi_\text{g} \\
\psi_\text{m}
\end{array} \right)
\end{displaymath}
and $\mathbf{V}$ is a $2\times 2$ matrix with components ~\eqref{V11},~\eqref{V22},~\eqref{V12ok}, and \eqref{V21ok}. The physical boundary conditions for the QNM problem are pure ingoing waves at the event horizon
\be
Y_n\sim b_n e^{-i\omega r_*},~~r_*\rightarrow -\infty,
\ee
and pure outgoing waves at spatial infinity
\be
Y_n\sim B_n e^{i\omega r_*},~~r_*\rightarrow +\infty,
\ee
where $Y_n$ is the $n$-th component of $\mathbf{Y}$, $b_n$ and $B_n$ are coefficients of the boundary conditions. With these boundary conditions, solving the QNFs is an eigenvalue problem.

The continued fraction method was first applied to gravitational problems by Leaver~\cite{Leaver1985}, and it has been used in coupled system~\cite{simmendinger1999,Rosa:2011my}. In order to get a recurrence relation, we need a suitable ansatz of the eigenfunction. Here, we assume that eigenfunctions of $\psi_\text{g}$ and $\psi_\text{m}$ are
\begin{widetext}
\beq
\psi_\text{g}&=&(r-r_S)^{-p}(r-r_S+1)^{p} e^{i(r-r_S)\omega}(r-r_S+1)^{i(r_B/2+r_S)\omega}\sum_n a^g_nH(r)^n, \label{cnmpsig}\\
\psi_\text{m}&=&(r-r_S)^{-p}(r-r_S+1)^{p} e^{i(r-r_S)\omega}(r-r_S+1)^{i(r_B/2+r_S)\omega}f_B(r)^{3/4}\sum_n a^m_nH(r)^n, \label{cnmpsim}
\eeq
\end{widetext}
where $p=\fc{i r_S^{3/2}\omega}{\sqrt{r_s-r_B}}$ and $H(r)=\fc{r-r_S}{r-r_B}$. Inserting these in to the master equations~\eqref{mastereq1} and ~\eqref{mastereq2}, we obtain seven-term recurrence relations
\beq
\mathbf{\alpha_0}\mathbf{A_1}&+&\mathbf{\beta_0}\mathbf{A_0}=0,\label{relation1}\\
\mathbf{\alpha_1}\mathbf{A_2}&+&\mathbf{\beta_1}\mathbf{A_1}+\mathbf{\gamma_1}\mathbf{A_0}=0,\label{relation2}\\
\mathbf{\alpha_2}\mathbf{A_3}&+&\mathbf{\beta_2}\mathbf{A_2}+\mathbf{\gamma_2}\mathbf{A_1}+\mathbf{\rho_2}\mathbf{A_0}=0,\label{relation3}\\
\mathbf{\alpha_3}\mathbf{A_4}&+&\mathbf{\beta_3}\mathbf{A_3}+\mathbf{\gamma_3}\mathbf{A_2}+\mathbf{\rho_3}\mathbf{A_1}+\mathbf{\lambda_3}\mathbf{A_0}=0,\label{relation4}\\
\mathbf{\alpha_4}\mathbf{A_5}&+&\mathbf{\beta_4}\mathbf{A_4}+\mathbf{\gamma_4}\mathbf{A_3}+\mathbf{\rho_4}\mathbf{A_2}+\mathbf{\lambda_4}\mathbf{A_1}\nn\\
&+&\mathbf{\sigma_4}\mathbf{A_0}=0,\label{relation5}\\
\mathbf{\alpha_n}\mathbf{A_{n+1}}&+&\mathbf{\beta_n}\mathbf{A_n}+\mathbf{\gamma_n}\mathbf{A_{n-1}}+\mathbf{\rho_n}\mathbf{A_{n-2}}+\mathbf{\lambda_n}\mathbf{A_{n-3}}\nn\\
&+&\mathbf{\sigma_n}\mathbf{A_{n-4}}+\mathbf{\delta_n}\mathbf{A_{n-5}}=0\label{relation6},
\eeq
where \begin{displaymath}
\mathbf{A_n}=
\left( \begin{array}{c}
a^g_n \\
a^m_n
\end{array} \right)
\end{displaymath}
is the vectorial coefficient. The coefficient matrices of the recurrence relations are very complicated, so we do not show the explicit expressions~\footnote{If you want the explicit expressions of the recurrence relations, please contact Wen-Di Guo through guowd@lzu.edu.cn}. Usually, a three-term recurrence relation can be obtained through a matrix-valued version of the Gaussian elimination~\cite{Leaver:1990zz,Berti:2009kk}. Then a matrix valued continued fraction can be solved and can be used to solve the QNFs. More details can be seen in Ref.~\cite{Pani:2013pma}. However, it needs to solve the inverse of the coefficient matrices of the recurrence relations again and again, which is difficult. There is an equivalent way to solve the QNFs. Equations~\eqref{relation1}-\eqref{relation6} can be written as
\begin{widetext}
\begin{displaymath}
\left( \begin{array}{cccccccccccccccc}
\beta^{11}_0&\beta^{12}_0&\alpha^{11}_0&\alpha^{12}_0&&&&&&&&&&&&\\
\beta^{21}_0&\beta^{22}_0&\alpha^{21}_0&\alpha^{22}_0&&&&&&&&&&&&\\
\gamma^{11}_1&\gamma^{12}_1&\beta^{11}_1&\beta^{12}_1&\alpha^{11}_1&\alpha^{12}_1&&&&&&&&&&\\
\gamma^{21}_1&\gamma^{22}_1&\beta^{21}_1&\beta^{22}_1&\alpha^{21}_1&\alpha^{22}_1&&&&&&&&&&\\
\rho^{11}_2&\rho^{12}_2&\gamma^{11}_2&\gamma^{12}_2&\beta^{11}_2&\beta^{12}_2&\alpha^{11}_2&\alpha^{12}_2&&&&&&&&\\
\rho^{21}_2&\rho^{22}_2&\gamma^{21}_2&\gamma^{22}_2&\beta^{21}_2&\beta^{22}_2&\alpha^{21}_2&\alpha^{22}_2&&&&&&&&\\
\lambda^{11}_3&\lambda^{12}_3&\rho^{11}_3&\rho^{12}_3&\gamma^{11}_3&\gamma^{12}_3&\beta^{11}_3&\beta^{12}_3&\alpha^{11}_3&\alpha^{12}_3&&&&&&\\
\lambda^{21}_3&\lambda^{22}_3&\rho^{21}_3&\rho^{22}_3&\gamma^{21}_3&\gamma^{22}_3&\beta^{21}_3&\beta^{22}_3&\alpha^{21}_3&\alpha^{22}_3&&&&&&\\
\sigma^{11}_4&\sigma^{12}_{4}&\lambda^{11}_4&\lambda^{12}_4&\rho^{11}_4&\rho^{12}_4&\gamma^{11}_4&\gamma^{12}_4&\beta^{11}_4&\beta^{12}_4&\alpha^{11}_4&\alpha^{12}_4&&&&\\
\sigma^{21}_4&\sigma^{22}_{4}&\lambda^{21}_4&\lambda^{22}_4&\rho^{21}_4&\rho^{22}_4&\gamma^{21}_4&\gamma^{22}_4&\beta^{21}_4&\beta^{22}_4&\alpha^{21}_4&\alpha^{22}_4&&&&\\
\delta^{11}_5&\delta^{12}_5&\sigma^{11}_5&\sigma^{12}_{5}&\lambda^{11}_5&\lambda^{12}_5&\rho^{11}_5&\rho^{12}_5&\gamma^{11}_5&\gamma^{12}_5&\beta^{11}_5&\beta^{12}_5&\alpha^{11}_5&\alpha^{12}_5&&\\
\delta^{21}_5&\delta^{22}_5&\sigma^{21}_5&\sigma^{22}_{5}&\lambda^{21}_5&\lambda^{22}_5&\rho^{21}_5&\rho^{22}_5&\gamma^{21}_5&\gamma^{22}_5&\beta^{21}_5&\beta^{22}_5&\alpha^{21}_5&\alpha^{22}_5&&\\
&\ddots&\ddots&\ddots&\ddots&\ddots&\ddots&\ddots&\ddots&\ddots&\ddots&\ddots&\ddots&\ddots&\ddots&\\
&&\delta^{11}_n&\delta^{12}_n&\sigma^{11}_n&\sigma^{12}_{n}&\lambda^{11}_n&\lambda^{12}_n&\rho^{11}_n&\rho^{12}_n&\gamma^{11}_n&\gamma^{12}_n&\beta^{11}_n&\beta^{12}_n&\alpha^{11}_n&\alpha^{12}_n\\
&&\delta^{21}_n&\delta^{22}_n&\sigma^{21}_n&\sigma^{22}_{n}&\lambda^{21}_n&\lambda^{22}_n&\rho^{21}_n&\rho^{22}_n&\gamma^{21}_n&\gamma^{22}_n&\beta^{21}_n&\beta^{22}_n&\alpha^{21}_n&\alpha^{22}_n
\end{array} \right)
\left(\begin{array}{c}
a^g_0\\
a^m_0\\
a^g_1\\
a^m_1\\
a^g_2\\
a^m_2\\
a^g_3\\
a^m_3\\
a^g_4\\
a^m_4\\
a^g_5\\
a^m_5\\
\vdots\\
a^g_n\\
a^m_n
\end{array}\right)=0.
\end{displaymath}
\end{widetext}
The QNFs are those which make the determinant of the coefficient matrix is zero. This method was first used to solve the QNFs of the RN black holes by Leaver~\cite{Leaver:1990zz}.

Except for the matrix-valued continued fraction method, we also use the matrix-valued direct integration method to solve the QNFs. More details can be seen in Ref.~\cite{Pani:2013pma}.

\begin{widetext}
\begin{center}
\begin{table}[!htb]
\begin{tabular}{|c|c|c|c|c|}
\hline
$Q_\text{m}/M$               &~Charged~BH~DI         &~Charged~BH~CF  &       $Q/M$~     &~~~~RN~BH    \\
\hline
    ~~                   &   $\omega_\text{R} M$~~~~$\omega_\text{I} M$~&   $\omega_\text{R} M$~~~~$\omega_\text{I} M$~& &$\omega_\text{R} M$~~~~~$\omega_\text{I} M$~\\
\hline
    0                    &            0.37367~~~~-0.088962   &            0.37367~~~~-0.088962  &0&    0.37367~~~~-0.088962\\
\hline
   0.2                   &            0.37474~~~~-0.089081   &            0.37480~~~~-0.089095  &0.2&    0.37474~~~~-0.089075\\
\hline
   0.4                   &            0.37848~~~~-0.089429   &            0.37855~~~~-0.089463  &0.4&   0.37844~~~~-0.089398\\
\hline
   0.6                   &            0.38641~~~~-0.089982   &            0.38649~~~~-0.090086   &0.6&  0.38622~~~~-0.089814\\
\hline
   0.8                   &            0.40163~~~~-0.090500   &            0.40169~~~~-0.090886  &0.8& 0.40122~~~~-0.089643\\
\hline
    1.12                 &            0.47027~~~~-0.084231   &            0.47153~~~~-0.092731     &0.9999& 0.43134~\cite{Onozawa:1995vu}~~~~-0.083460~\cite{Onozawa:1995vu}\\
\hline

\end{tabular}\\
\caption{The fundamental QNMs for the gravitational field $\psi_\text{g}$ of the charged black hole with scalar hair (using the direct integration (DI) method and using the continued fraction (CF) method) and the RN black hole for different values of the magnetic charge $Q_\text{m}$ and electric charge $Q$, respectively. The angular number $l$ is set to $l=2$.}
\label{realpart}
\end{table}
\end{center}

\begin{center}
\begin{table}[!htb]
\begin{tabular}{|c|c|c|c|c|}
\hline
$Q_\text{m}/M$               &~Charged~BH~DI       &~Charged~BH~CF &       $Q/M$~     &~~~~RN~BH    \\
\hline
    ~~                   &   $\omega_\text{R} M$~~~~$\omega_\text{I} M$~&   $\omega_\text{R} M$~~~~$\omega_\text{I} M$~& &$\omega_\text{R} M$~~~~~$\omega_\text{I} M$~\\
\hline
    0                    &            0.45715~~~~-0.094784    &            0.45715~~~~-0.094784 &0&    0.45759~~~~-0.095004\\
\hline
   0.2                   &            0.46295~~~~-0.095377    &            0.46296~~~~-0.095359 &0.2&    0.46297~~~~-0.095373\\
\hline
   0.4                   &            0.47969~~~~-0.096462    &            0.47969~~~~-0.096441 &0.4&   0.47993~~~~-0.096442\\
\hline
   0.6                   &            0.51053~~~~-0.098155    &            0.51055~~~~-0.098133 &0.6&   0.51201~~~~-0.098017\\
\hline
   0.8                   &            0.56316~~~~ -0.10008    &            0.56320~~~~ -0.10002 &0.8&   0.57013~~~~-0.099069\\
\hline
    1.12                   &            0.78258~~~~-0.091135  &            0.79925~~~~-0.098085   &0.9999& 0.70430~\cite{Onozawa:1995vu}~~~~-0.085973~\cite{Onozawa:1995vu}\\
\hline

\end{tabular}\\
\caption{The fundamental QNMs for the magnetic field $\psi_\text{m}$ of the charged black hole with scalar hair (using the direct integration (DI) method and using the continued fraction (CF) method) and the electric field $\psi_\text{e}$ of the RN black hole for different values of the magnetic charge $Q_\text{m}$ and electric charge $Q$, respectively. The angular number $l$ is set to $l=2$.}
\label{impart}
\end{table}
\end{center}
\end{widetext}

\begin{figure*}[htb]
\begin{center}
\subfigure[~The real parts of the QNFs for the charged black hole with scalar hair.]  {\label{TSreals}
\includegraphics[width=7cm]{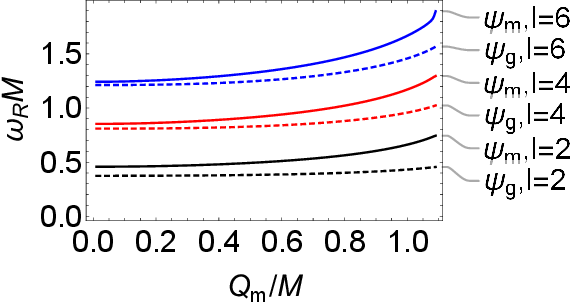}}
\subfigure[~The real parts of the QNFs for the RN black hole.]  {\label{RNreals}
\includegraphics[width=7cm]{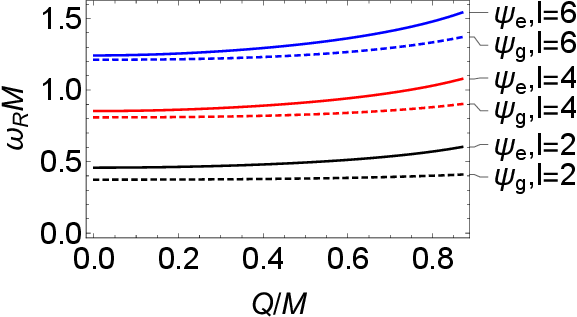}}
\end{center}
\caption{The effects of the magnetic charge $Q_\text{m}$ of the charged black hole with scalar hair and the electric charge $Q$ of the RN black hole on the real parts of the fundamental QNFs. The solid and dashed lines correspond to the QNFs of the magnetic field $\psi_\text{m}$ (or the electric field $\psi_\text{e}$) and the gravitational field $\psi_\text{g}$, respectively. The black, red, and blue lines correspond to the QNFs with $l=2$, $l=4$, and $l=6$, respectively. (a) The real parts of the QNFs for the charged black hole with scalar hair. (b) The real parts of the QNFs for the RN black hole.}
\label{TSRN}
\end{figure*}

\begin{figure*}[htb]
\begin{center}
\subfigure[~The imaginary parts of the QNFs for the charged black hole with scalar hair.]  {\label{TSimss}
\includegraphics[width=7cm]{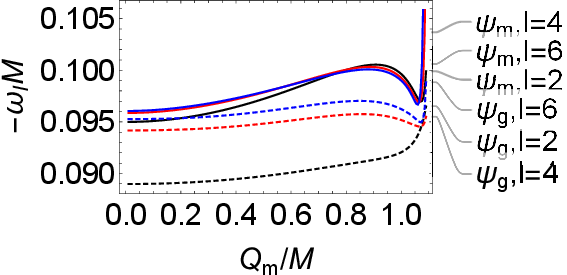}}
\subfigure[~The imaginary parts of the QNFs for the RN black hole.]  {\label{RNims}
\includegraphics[width=7cm]{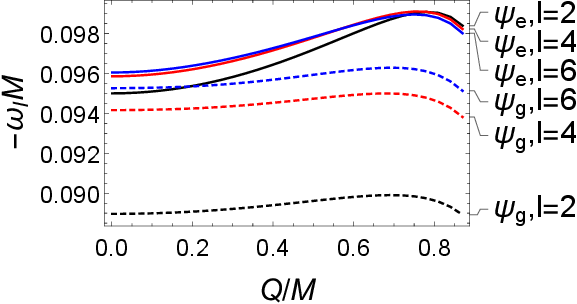}}
\end{center}
\caption{The effects of the magnetic charge $Q_\text{m}$ of the charged black hole with scalar hair and the electric charge $Q$ of the RN black hole on the imaginary parts of the fundamental QNFs. The solid and dashed lines correspond to the QNFs of the magnetic field $\psi_\text{m}$ (or the electric field $\psi_\text{e}$) and the gravitational field $\psi_\text{g}$, respectively. The black, red, and blue lines correspond to the QNFs with $l=2$, $l=4$, and $l=6$, respectively. (a) The imaginary parts of the QNFs for the charged black hole with scalar hair. (b) The imaginary parts of the QNFs for the RN black hole.}
\label{TSims}
\end{figure*}

We solve the fundamental QNMs numerically, which dominate the ringdown waveform at late time. The values of the frequencies of fundamental QNMs for the gravitational field $\psi_\text{g}$ and the magnetic field $\psi_\text{m}$ for different values of the magnetic charge $Q_\text{m}$ with $l=2$ are shown in Tables~\ref{realpart} and \ref{impart}. When $Q_\text{m}=0$, the metric~\eqref{metric_four} reduces to the Schwarzschild metric. The master equation ~\eqref{mastereq1} reduces to the odd parity gravitational perturbation of the Schwarzschild black hole in general relativity. The QNFs are also the same as the Schwarzschild black hole case. This confirms that our numerical method is valid. Besides, the QNFs solved by the matrix-valued direct integration method and the matrix-valued continued fraction method agree well each other, which can be seen in Tables~ \ref{realpart} and \ref{impart}. This strengthens the validity of our results. Note that the charge of the charged black hole with scalar hair can be seen as a dark charge. One of the effects of this charge is to stabilize the spacetime. Besides, it is possible that microscopic topology stars could be candidates for dark matter. In this paper, we would like to compare our results with that of the RN black hole. Comparing the QNFs of the charged black hole with scalar hair and the RN black hole, we can see that, the differences of their numerical values are very small. So we almost can not distinguish them from the gravitational wave data. Note that, for the extreme RN black hole, the singular structure of the perturbation equations is different from the nonextreme ones. The QNMs for the maximally charged RN black hole were studied in Ref.~\cite{Onozawa:1995vu}. Our results for the RN black hole with $Q/M=0.9999$ are taken from that paper. It is valuable to compare the QNFs of nearly extremal charged black hole with that of the RN black hole. However, we can only calculate the QNFs for $Q_\text{m}/M=1.12$, or equivalently, $  Q_\text{m}/M = 0.96995\times (2/\sqrt{3})$. More extremal case needs special concern. 

The effects of the magnetic charge $Q_\text{m}$ of the charged black hole with scalar hair and the electric charge $Q$  of the RN black hole on the fundamental QNMs are shown in Figs.~\ref{TSRN} and~\ref{TSims}. From Figs.~\ref{TSreals} and \ref{RNreals}, it can be seen that the real parts of the QNFs for both black holes increase with the magnetic charge $Q_\text{m}$ or the electric charge $Q$. The imaginary parts of the QNFs for the RN black hole first increase, then decrease as the electric charge $Q$ increases, which can be seen in Fig.~\ref{RNims}. However, the situation for the imaginary parts of the charged black hole with scalar hair is different, which can be seen in Fig. \ref{TSims}. The imaginary part for the gravitational field $\psi_\text{g}$ of the charged black hole with scalar hair when $l=2$ (the black dashed line in Fig.~\ref{TSimss}) increases with the magnetic charge $Q_\text{m}$. The imaginary part for the gravitational field $\psi_\text{g}$ when $l>2$ and for the magnetic field $\psi_\text{m}$ when $l\geq2$ first increases, then decreases, and finally increases as the magnetic charge $Q_\text{m}$ increases. 
%When $2\leq l\leq 6$, an interesting phenomenon appears. \textcolor[rgb]{0.00,0.07,1.00}{There are several peaks of the \textcolor[rgb]{0.00,0.07,1.00}{QNFs'} imaginary parts for both the magnetic field $\psi_\text{m}$ (solid lines in Fig.~\ref{TSims}) and the gravitational field $\psi_\text{g}$ (dashed lines in Fig.~\ref{TSims}). And the number of peaks of \textcolor[rgb]{0.00,0.07,1.00}{QNFs'} imaginary parts for the gravitational field $\psi_\text{g}$ is always less than that of the magnetic field $\psi_\text{m}$.} This phenomenon does not appear in the RN black hole. When $l>6$, we do not find such particular phenomenon anymore. We can see from Fig.~\ref{TSl7} that, when $l=7$ and $l=8$, the imaginary parts for both the magnetic field $\psi_\text{m}$ and the gravitational field $\psi_\text{g}$ increase with the magnetic charge $Q_\text{m}$. When $l=9$ and $l=10$, the imaginary parts for the magnetic field $\psi_\text{m}$ decrease first then increase as the magnetic charge $Q_\text{m}$ increases, while the imaginary parts for the gravitational field $\psi_\text{g}$ increase with the magnetic charge $Q_\text{m}$.

\section{Conclusions}\label{conclusion}

In five-dimensional spacetime, based on the Einstein-Maxwell action~\eqref{action5}, Bah and Heidmann proposed a nonsingular black hole/topology star. This is similar to the classical black hole in macrostate geometries, more importantly, it can be constructed from type IIB string theory. Integrating the extra dimension $y$, the five-dimensional Einstein-Maxwell theory reduces to a four-dimensional Einstein-Maxwell-dilaton theory which supports a spherically static black hole/topological star solution with a magnetic charge.

We investigated the QNMs of the charged black hole with scalar hair by studying the linear perturbation of the gravitational field and the electromagnetic field. Because of the spherical symmetry of the background spacetime, the radial parts of the perturbed fields can be decomposed from the angular parts. The angular parts can be expanded by the spherical harmonics. The background scalar field~\eqref{solutionphi} and metric field~\eqref{metric_four} are even parity under the space inversion, however, the background magnetic field~\eqref{solutionnF} is odd parity. So the scalar perturbation and  even-parity parts of the metric perturbations couple to the odd-parity parts of the electromagnetic perturbations to the linear order, and the odd-parity parts of the metric perturbations couple to the even-parity parts of the electromagnetic perturbations to the linear order, which we named as type-I and type-II couplings, respectively. For simplicity, we study the type-II coupling perturbations. Finally, we obtained two coupled perturbation equations~\eqref{mastereq1} and \eqref{mastereq2}. Although the extra dimension radius $R_y$ can be eliminated from the master equations by a transformation of the electromagnetic field $\psi_\text{m}$, it can also affect the QNM spectrum through the gravitational constant.

Using the matrix-valued direct integration method and the matrix-valued continued fraction method, we computed the fundamental QNFs for both the gravitational perturbation and the magnetic field perturbation, which will dominate the ringdown wave at late time. The values of the frequencies of the fundamental QNMs for the gravitational field $\psi_\text{g}$ and the magnetic field $\psi_\text{m}$ for different values of the magnetic charge $Q_\text{m}$ with $l=2$ are shown in Tables~\ref{realpart} and \ref{impart}. The results obtained from the matrix-valued direct integration method and the matrix-valued continued fraction method agree well each other, which strengthens the validity of our results. The differences of the frequencies of the fundamental QNMs between the charged black hole with scalar hair and the RN black hole are very small. So we almost can not distinguish them from the gravitational wave data. The effect of the magnetic charge $Q_\text{m}$ of the charged black hole with scalar hair on the fundamental QNFs are shown in Figs.~\ref{TSreals},~\ref{TSims}. The real parts of the QNFs increase with the magnetic charge $Q_\text{m}$, which is similar to that of the RN black hole. However, the situation for the imaginary parts of the QNFs of the charged black hole with scalar hair is different, which can be seen in Fig.~\ref{TSims}.

We only studied the type-II coupling perturbations, where the scalar field does not couple to the other two fields. So we expect that the type-I coupling perturbations will give us more information about the charged black hole with scalar hair, which will be studied in the future.

\section{Acknowledgments}

We thank Pierre Heidmann for important comment, suggestion and discussion. This work was supported by National Key Research and Development Program of China (Grant No. 2020YFC2201503), the National Natural Science Foundation of China (Grants No. 12205129, No. 12147166, No. 11875151, No. 12075103, and No. 12247101), the China Postdoctoral Science Foundation (Grant No. 2021M701529), the 111 Project (Grant No. B20063), and Lanzhou City's scientific research funding subsidy to Lanzhou University.

\begin{widetext}
\begin{appendices}
\section{Explicit perturbation equations}\label{appendix}

In this appendix we give the details of how to get the master perturbation equations ~\eqref{mastereq1} and \eqref{mastereq2}. The nonvanishing parts of the perturbed Einstein equations are the $(t,\phi)$, $(r,\phi)$, and $(\theta,\phi)$ components
\beq
&&2 e \Big(4 f_B  \fc{r_S}{r}-f_S\fc{ r_B }{ r}-\fc{f_S r_B^2}{f_B r^2}  +
  4 l(l+1)  + 10 f_S  \fc{r_B}{r} + 8  \fc{r_S}{r}
+ 12 \fc{r_B r_S}{r^2}\Big)h_0-8ef_Bf_S r^2 h_0''\nn\\
&&-4 ie f_S r \omega  \left(r f_B'+4 f_B\right)h_1
-8ief_Bf_S r^2 \omega h_1'=-16a\sqrt{G_4}\sqrt{f_B}f_{02}\label{eqtphi},
\eeq
\beq
&&8 e f_B^2 \left(r^4 \omega ^2-f_S \left(r_S (2 f_B+3 r_B)-2 r f_B  r_S+(l(l+1)-2) r^2\right)\right)h_1+4i\fc{e}{r}f_B\omega(4f_B+rf_B')h_0\nn\\
&&-8ier^4f_B^2\omega h_0'=16a\sqrt{G_4}r^2f_B^{5/2}f_Sf_{12}\label{eqrphi},
\eeq
\beq
2f_S  h_1 f_B f_S'+f_S^2 \left(h_1 f_B'+2 f_B h_1'\right)+2 i h_0 \omega=0\label{eqthetaphi},
\eeq
where the constant $a$ is defined as $a\equiv e\sqrt{3r_B r_S}$. And the nonvanishing parts of the perturbed Maxwell equations are the $t$, $r$, and $\theta$ components
\beq
f_S r\left( r f_B'+4  f_{B}\right)f_{01}+2 f_B f_S r^2 f_{01}'-2l(l+1)f_{02}&=&\fc{a}{f_B r^2 \sqrt{G_4}}l(l+1)h_0,\label{eqat}\\
2i\omega \sqrt{f_B}r^4 f_{01}+2f_S\sqrt{f_B}r^2l(l+1)f_{12}&=&-\fc{a}{\sqrt{G_4}} f_S l(l+1)h_1,\label{eqar}\\
2 f_B^{3/2} f_S \kappa_4 r^3 (f_{12} f_S)'+\sqrt{f_B} r^3 \left(f_{12} f_S^2 f_B'+2 i f_{02} \omega \right)&=&\frac{a}{\sqrt{G_4}} \left(3 f_S h_1-f_B f_S (f_S  r h_1)'- \omega r h_0\right).\label{eqatheta}
\eeq
Actually, among the six perturbed equations only four of them are independent. Equation~\eqref{eqtphi} can be derived from Eqs.~\eqref{eqrphi},~\eqref{eqthetaphi}, and \eqref{eqatheta} with the background Einstein equation~\eqref{EFg}. Similarly, Eq.~\eqref{eqatheta} can also be obtained by using Eqs.~\eqref{eqar},~\eqref{eqat}. \textcolor[rgb]{0.00,0.07,1.00}{Therefore}, we can use four independent equations~\eqref{eqrphi}-\eqref{eqar} and an identity~\eqref{eqf01} to solve five independent variables $h_0$, $h_1$, $f_{01}$, $f_{02}$, and $f_{12}$.

The variable $h_0$ can be solved from Eq.~\eqref{eqthetaphi} as
\beq
h_0=\fc{i}{2\omega}f_S\lt(f_S f_B'+2f_B f_S'\rt)h_1+2f_S^2 f_B h_1'.\label{eqh0}
\eeq
Using this formula and Eqs.~\eqref{eqat} and \eqref{eqar}, we can obtain $f_{02}$ and $f_{12}$ in terms of $h_1$ and $f_{01}$ as
\beq
f_{02}&=&\fc{f_S r}{2l(l+1)}\lt[2rf_B f_{01}'+(4f_B+r f_B')f_{01}\rt]-\fc{i a f_S}{4r^2\sqrt{G_4}\omega \sqrt{f_B}}\lt[2f_B(h_1 f_S)'+f_S f_B' h_1\rt],\label{eqf02}\\
f_{12}&=&-\fc{i\omega r^2}{f_S (l+1)l}f_{01}-\fc{a}{2\sqrt{G_4}\sqrt{f_B}r^2}h_1\label{eqf12}.
\eeq
Substituting Eqs.~\eqref{eqh0}-\eqref{eqf12} into Eqs.~\eqref{eqrphi} and \eqref{eqf01} we can obtain two second order differential equations in which $h_1$ and $f_{01}$ are coupled
\beq
&-&\fc{1}{2}\sqrt{f_B}f_S h_1''+\lt[\fc{\sqrt{f_B}}{2r^2}(2rf_S-3r_S)-\fc{r_B f_S}{2r^2\sqrt{f_B}}\rt]h_1'+\Big[\fc{r_B^2f_S}{8r^4f_B^{3/2}}-\fc{\omega^2}{2\sqrt{f_B}f_S}\nn\\
&-&\fc{1}{4r^4\sqrt{f_B}}\big((3r_Br_S-2(l-1)(l+2)r^2)-5r r_B f_S\big)+\fc{\sqrt{f_B}}{2r^4 f_S}(4r r_S f_S-r_S^2)\Big]h_1=\fc{2i \omega a \sqrt{G_4}}{e^2l(l+1)f_S}f_{01},\label{eqh1f01}
\eeq
\beq
&-&\fc{r^2f_Sf_B }{l(l+1)}f_{01}''+\fc{2f_B(r_S+4rf_S)-3r_B f_S}{2l(l+1)}f_{01}'+ \lt[1-\fc{r^2\omega^2}{l(l+1)f_S}+\fc{f_S'(4rf_B+r_B)}{2l(l+1)}-\fc{f_S(5r_B+4rf_Bf_S)}{2l(l+1)r}\rt]f_{01}\nn\\
&=&-\fc{ia\sqrt{f_B}f_S^2}{2r^2\omega \sqrt{G_4}}h_1''-\fc{ia f_S[r_B f_S+f_B(3r_S-2rf_S)]}{2r^4\omega \sqrt{G_4} f_B}h_1'-\fc{ia r_B^2f_S^2}{8r^6\omega\sqrt{G_4} f_B^{3/2}}h_1\nn\\
&+&\lt[\fc{i a r_S\sqrt{f_B}(r_S-3r f_S)}{2r^6\omega\sqrt{G_4}}-\fc{i a (3r r_B f_S^2-2r^4\omega^2-3r_B r_S f_S)}{4r^6\omega\sqrt{G_4}\sqrt{f_B}}\rt]h_1.\label{eqf01h1}
\eeq
In order to get the Schr\"{o}dinger-like form, we need to define the following master variables
\beq
\psi_\text{g}\equiv f_B^{1/4}f_S\fc{1}{r}h_1,\\
\psi_\text{m}\equiv \sqrt{f_B}r^2f_{01}.
\eeq
In the tortoise coordinate $r_*$, Eqs.~\eqref{eqh1f01} and \eqref{eqf01h1} can be rewritten into the form of Eqs.~\eqref{mastereq1} and \eqref{mastereq2}.

\end{appendices}
\end{widetext}
%\bibliographystyle{JHEP}
%
%\bibliography{D:/W_D_Guo/works/topology_star/shadow/referrenc/refs}
%\bibliography{E:/works/topology_star/shadow/referrenc/refs}
%\providecommand{\href}[2]{#2}\begingroup\raggedright\begin{thebibliography}{10}

%
%

\end{document}